\documentclass{PoS}

\title{On the calculation and use of non-zero momentum correlators in 
  lattice simulations.}

\ShortTitle{non-zero momentum correlators}

\author{\speaker{Claudio Rebbi}\\
        Boston University\\
        E-mail: \email{rebbi@bu.edu}}

\abstract{In lattice simulations one generally projects correlators over 
  zero spatial momentum to calculate masses and related spectral
  data. The sum over space lattice points, however, discards
  information which may be useful especially in the calculation of
  disconnected diagrams. By using momentum conservation, the
  calculation of non-zero momentum components of disconnected diagrams
  and other quantities related to space convolutions can be done with
  little additional computational cost and may be useful in the
  analysis of disconnected correlators.}

\FullConference{The 36th Annual International Symposium on Lattice Field 
  Theory - LATTICE2018\\22-28 July, 2018\\
  Michigan State University, East Lansing, Michigan, USA.}

\begin{document}

\section{Introduction}
\label{intro}

In lattice gauge theory simulations the calculation of correlators of
particle whose source and sink have non-zero vacuum expectation value
is challenging because of the need to include disconnected
contributions.  This happens, for example, in the calculation of
glueball correlators and, most notably, in the calculation of the
correlator of a flavor-singlet scalar particle.  In this case the
disconnected component involves a correlation of local operators:
\begin{equation}
  D(t)=(1/N_x N_t)\sum_{\vec x, \vec y, t'} \langle \phi(\vec x,t') 
  \phi(\vec y, t'+t) \rangle
  \label{eq1}
\end{equation}
where $\phi(\vec x,t')$ is a closed fermion propagator, i.e.~the trace
of a fermion propagator originating and ending at $\vec x,t'$.  In
Eq.~\ref{eq1}, as is commonly done, one sums over the space positions
$\vec x, \vec y$ of $\phi$, in order to project the correlator over
zero momentum states.  As a matter of fact a single sum, over $\vec x$
or $\vec y$, would in principle be sufficient to project over zero
momentum, but the correlation of the individual terms in Eq.~\ref{eq1}
is affected by large statistical fluctuations, which one reduces by
summing over both $\vec x$ and $\vec y$.  Still, for lattices of large
size, calculating all the individual closed propagators in
Eq.~\ref{eq1} over all the lattice configurations produced in the
simulation would be too time consuming, and one resorts to suitable
computational techniques, such as stochastic source methods, to
accelerate the calculation.  We will not need to be concerned here
with these techniques.  The only important fact for us is that, prior
to the sums over $\vec x$ and $\vec y$, they provide a reasonably
precise evaluation of the individual closed propagators at all points
of the lattice.

For simplicity from now on we will refer to the disconnected component
$D(t)$ of the correlator simply as the ``disconnected correlator''.
Correspondingly, we will use the term ``connected correlator'' to
indicate the connected component $C(t)$ of the correlators, where a
fermion propagator from $\vec x,t'$ to $\vec y, t'+t$ is combined with
a propagator from $\vec y, t'+t$ to $\vec x,t'$, with a suitable sum
over spin and flavor indices.  Also, we will leave normalization
factors implicit.

The main point of this note is that, for large separation, $\phi(\vec
x,t')$ and $\phi(\vec y, t'+t)$ in Eq.~\ref{eq1} are essentially
uncorrelated but still contribute to the noise.  As, with the advance
of computer power, simulations with lattices of increasingly larger
size become possible, the problem grows more severe. E.g.~with
$N_x=256$ for $t=32$ the separation between $\phi(\vec x,t')$ and
$\phi(\vec y, t'+t)$ ranges from $32$ to $\sqrt{32^2+3 \times 128^2} =
224$ lattice spacings.  We may then try to reduce the statistical
noise by giving higher weight to the operators facing each other, and
this will be our first consideration.  The above weighting procedure,
which involves a convolution, is most easily done in momentum space.
As we shall see, a byproduct is that one can easily calculate the
disconnected correlators for non vanishing momentum, and this will
presented in a subsequent section.

\section{Gaussian weighted correlators}
\label{gaussian}

We try to reduce the noise due to far away $\phi$ operators by
correlating $\phi(\vec x,t)$ with a distribution of $\phi(\vec y,t')$
centered around $\vec x$ rather then the whole time slice at $t'$.
The most obvious distribution is a Gaussian one, although one could
use different weight factors.  Thus we define a ``Gaussian weighted
correlator'' as
\begin{equation}
  D_G(t)=\sum_{\vec x, \vec y, t'} \langle \phi(\vec x,t') 
  G(\vec y) \phi(\vec x+\vec y, t'+t) \rangle
  \label{eq2}
\end{equation}
where $G(\vec y)$ is a Gaussian weight function falling off with
distance. This may reduce the statistical noise at the expense of the
introduction of non-zero momentum components.

Computing the above correlator in configuration space would be too
demanding, it would involve the averaging over $N_x^6$ products of
$\phi(\vec x,t')$ with $\phi(\vec x+\vec y, t'+t)$, but the
convolution in Eq.~\ref{eq2} can be much more readily computed in
momentum space. We express $\phi(\vec x, t)$ and $G(\vec x)$ in terms
of their (space) Fourier transforms, which we denote by a tilde (we
continue to neglect normalization factors in order to streamline the
formulae)
\begin{eqnarray}
  \phi(\vec x, t) &=& \sum_{\vec p} \tilde \phi(\vec p, t) 
  e^{\imath \vec p \cdot \vec x} \label{eq3a} \\
  G(\vec x) &=& \sum_{\vec p} \tilde G(\vec p)
  e^{-\imath \vec p \cdot \vec x} \label{eq3b}
\end{eqnarray}
and obtain
\begin{equation}
  D_G(t)=\sum_{\vec x, \vec y, \vec p, \vec q,\vec r,t'} 
  \tilde G(\vec q) \langle \tilde \phi^*(\vec p,t') 
  \tilde \phi(\vec r, t'+t) \rangle e^{\imath[- \vec q \cdot \vec y
    -\vec p \cdot \vec x + \vec r \cdot  (\vec x+ \vec y)]}
  \label{eq4}
\end{equation}
(where we used the reality of $\phi(\vec x, t)$.)  
\begin{figure}[t]
  \includegraphics[width=.49\textwidth]{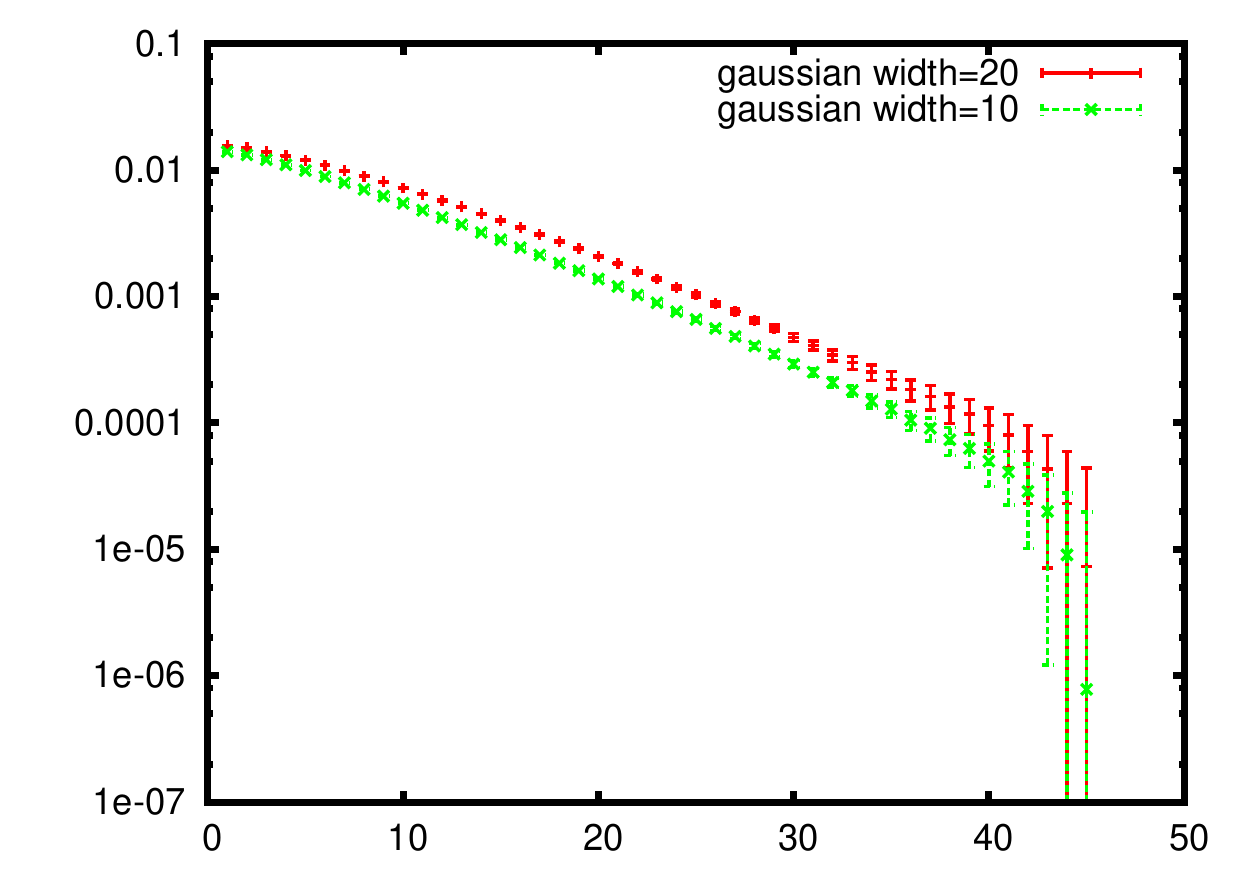}
  \includegraphics[width=.49\textwidth]{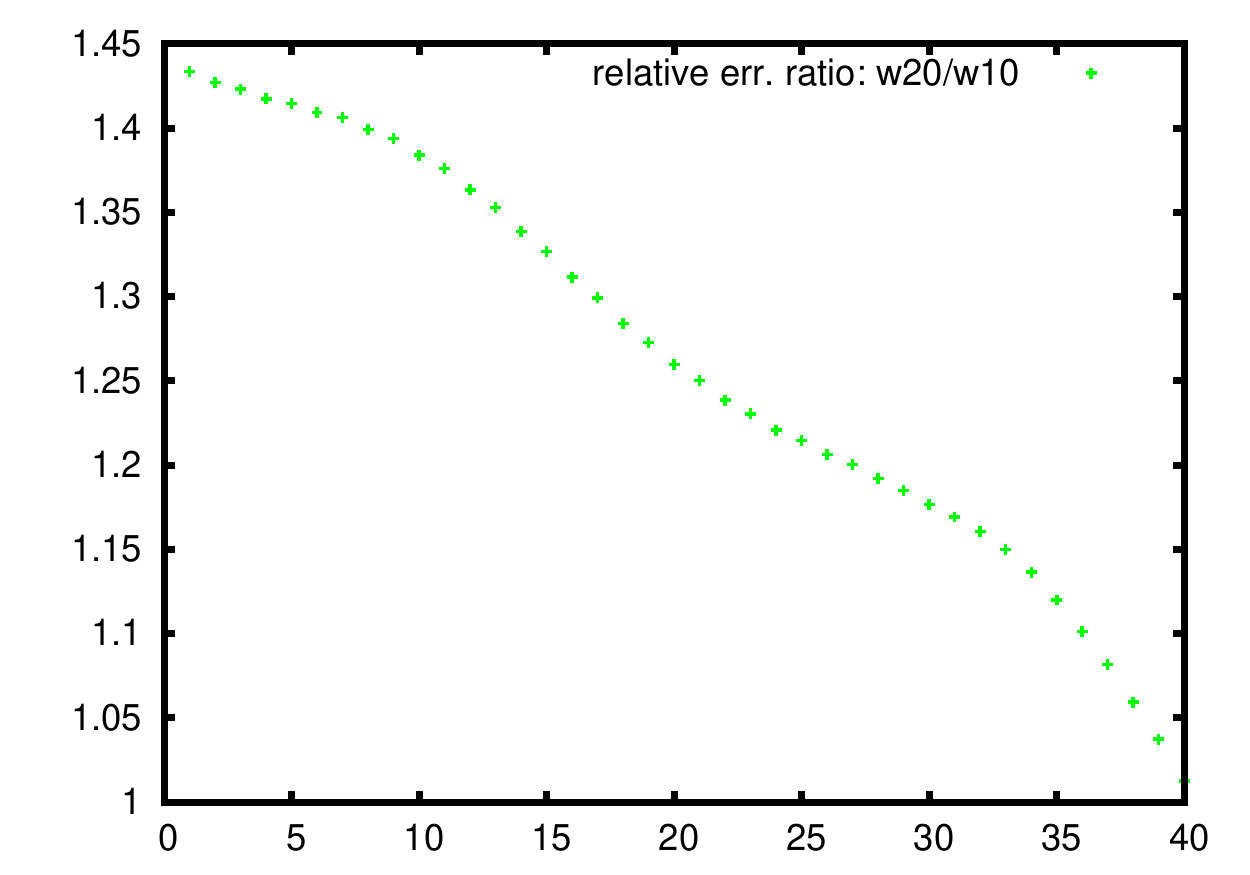}
  \caption{Left: The Gaussian weighted disconnected correlators discussed
    in the text for widths of 10 and 20 lattice spacings.  Right: the ratio
    of relative errors for width 20 over the relative errors for width 10.}
  \label{fig1}
\end{figure}
And now we make use of the crucial fact that correlators are diagonal
in momentum.  This property is not true configuration by
configuration, but is true of the exact correlators, and will be
better and better satisfied as one improves the statistics.  So we
assume it as true in Eq.~\ref{eq4}.  This introduces a delta function
forcing $\vec r= \vec p$ and giving
\begin{eqnarray}
  D_G(t)=\sum_{\vec x, \vec y, \vec p, \vec q,t'} 
  \tilde G(\vec q) \langle \tilde \phi^*(\vec p,t') 
  \tilde \phi(\vec p, t'+t) \rangle e^{\imath[-\vec q \cdot \vec y
    + \vec p \cdot \vec y)]}
  \nonumber \\
  =\sum_{\vec p, t'} \tilde G(\vec p) \langle \tilde \phi^*(\vec p,t') 
  \tilde \phi(\vec p, t'+t) \rangle
  \label{eq5}
\end{eqnarray}
From the computational point of view this equation leads to the
following procedure.  As one computes the closed fermion propagators
$\phi(\vec x, t)$ configuration by configuration one should calculate
right away their Fourier transforms $\tilde \phi(\vec p, t)$, or save
the values of $\phi(\vec x, t)$ at all points $\vec x$ (rather than
sum right away over $\vec x$) for a subsequent Fourier transform.  
The products of these Fourier transforms, averaged over all
configurations, are then summed over $\vec p$ with the Gaussian weight
factor $\tilde G(\tilde)$ to obtain the Gaussian weighted correlators
$D_G(t)$.  For the full correlator of the state under consideration
one should, of course, combine the disconnected correlator with the
connected one with the appropriate coefficients, but the calculation
of the Fourier transform of the connected correlator is
straightforward.

We tested the above procedure on data for a scalar singlet correlator
($\sigma$ particle) obtained by the LSD collaboration\footnote{The 
  Lattice Strong Dynamics collaboration: 
  Xiao-Yong Jin, James Osborn (ANL); 
  Richard Brower, Claudio Rebbi~(BU);
  Pavlos Vranas (LLNL); Evan Weinberg (NVIDIA); Enrico Rinaldi (RBRC);
  David Schaich (U.~Bern);
  Joseph Kiskis (UC~Davis); 
  Anna Hasenfratz, Ethan Neil (joint w/RBRC), Oliver Witzel (U.~Colorado);
  Graham Kribs (U.~Oregon); Thomas Appelquist,
  Kimmy Cushman, George Fleming, Andy Gasbarro (Yale).}
in the course of a study of the SU(3) gauge theory with eight
dynamical fermions~\cite{Appelquist:2018yqe}.  We refer
to~\cite{Appelquist:2018yqe} for the detailed parameters of the
simulation.  The data we are showing have been obtained from averages
over 2393 thermalized configurations, divided into three approximately
equally numbered streams, on a $64 \times 128$ lattice with
$\beta_F=4.8,\, \beta_A/\beta_F=-0.25$, and bare fermion mass $am=
0.00125$ in lattice units.  A straightforward implementation of the
Gaussian smearing would not be very revealing, because the
fluctuations in $D(t)$ and $D_G(t)$ are dominated by the fluctuations
of the vacuum expectation value of the disconnected correlators.
However, the computational procedure outlined above permits to get rid
very easily of the unwanted vacuum contribution.  One only needs to
exclude the value $\vec p=0$ from the sum in Eq.~\ref{eq5}.  In
Figure~\ref{fig1} we show the disconnected $\sigma$ correlator
weighted with Gaussians of widths $w=10$ and $w=20$ lattice units and
with zero-momentum subtracted.  With the large statistics the errors
for $t \lesssim 30$ are not very visible in the graph at left, making
it difficult to compare the relative accuracies of the two different
Gaussian smearing, but in the graph at right we show the ratio of the
relative errors in the data obtained with $w=20$ over those obtained
with $w=10$.  The results confirms that, as we would have expected,
giving greater weight to the operators closer in space reduces the
error due to statistical noise.  As the separation in time increases
the advantage gradually disappears, as could have also been expected,
since for larger times the separation in time weighs more than the
separation in space.

\section{Momentum dependent correlators}
\label{momentum}

\begin{figure} [h!]
  \begin{center}
    \includegraphics[width=.6\textwidth]{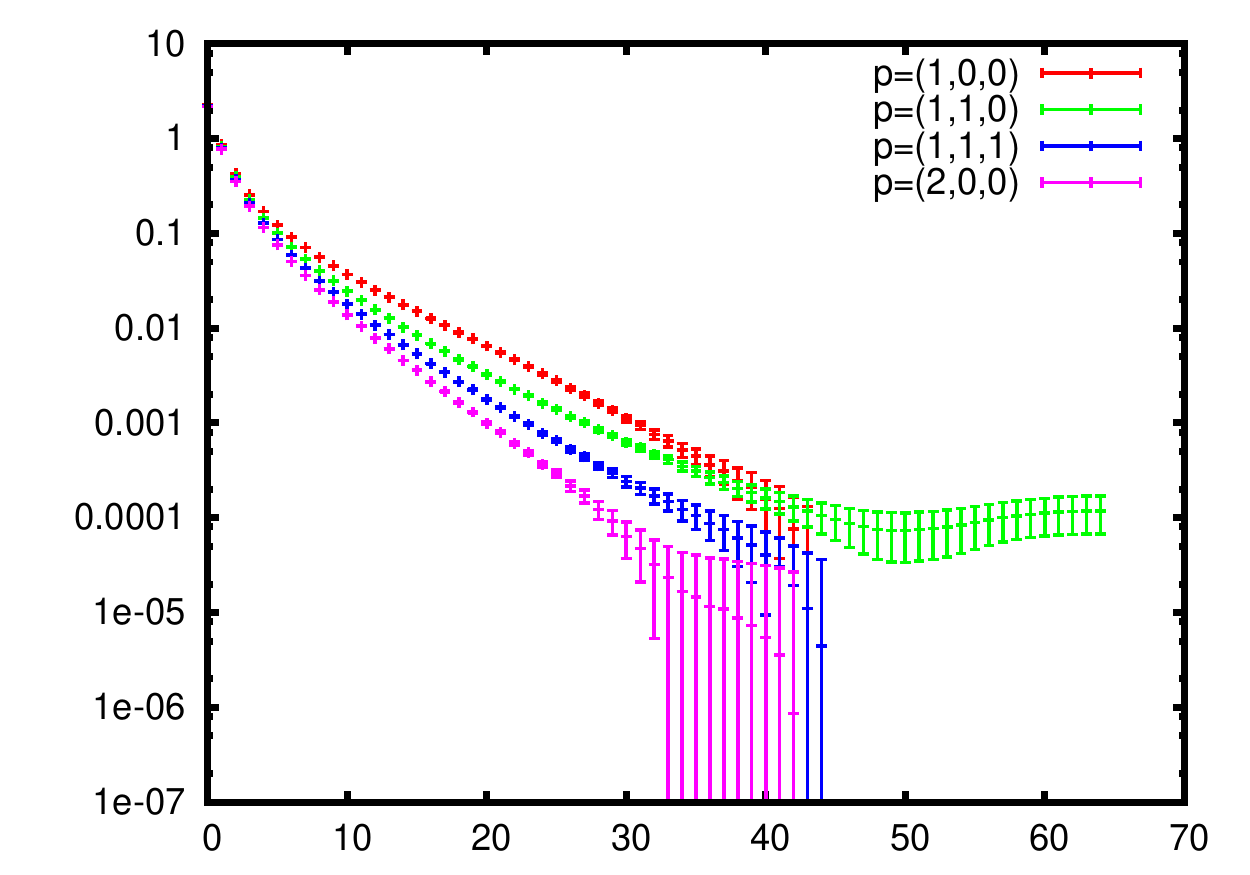}
    \caption{The full $\sigma$ correlators for the four lowest 
      non-zero momenta from the simulations on a $64\times 128$
      lattice, $m=0.00125$, mentioned in the text.}
    \label{fig2}
  \end{center}
\end{figure}

A byproduct of the procedure described above is that it can also be
used to easily calculate the individual momentum components of the
disconnected correlators.  Always under the assumption that momentum
is conserved, which will be better and better satisfied as the
statistics increases, the momentum components of the disconnected
correlators are calculated by taking first the spatial Fourier
transforms of the the closed fermion propagators as they are
calculated configuration by configuration.  Then the expectation value
of the products of the Fourier transforms (FT) (more precisely of the
FT and its conjugate) at definite time separation are averaged over
all thermalized configurations.  The disconnected correlators with at
least one non-zero momentum component have no vacuum expectation value
and including them in the overall analysis of the data may help
improve the accuracy of the singlet scalar mass determination.  Of
course, the exponential behavior of the correlator (with disconnected
and connected components combined) will provide a determination of the
energy of the state rather than of its mass, so if the value of the
lowest momentum is comparable or larger than the mass one wishes to
determine, it will become more difficult to extract a reliable value
for the mass from the correlators with non-vanishing momentum.  Thus,
as in the case of the Gaussian weighted correlators, the use of
non-zero momentum components will become more valuable as the size 
of the lattices that can be simulated will increase.

In Figure~\ref{fig2} we show the four lowest non-zero momentum
components of the full correlator $2*D(t)-C(t)$, $C(t)$ being the 
connected correlator, of the $\sigma$ (scalar singlet) state
in the analysis of the 2393 thermalized configurations, with  
lattice size $64 \times 128$ and bare fermion mass $am= 0.00125$,
referred to above.

\begin{figure}
  \includegraphics[width=.49\textwidth]{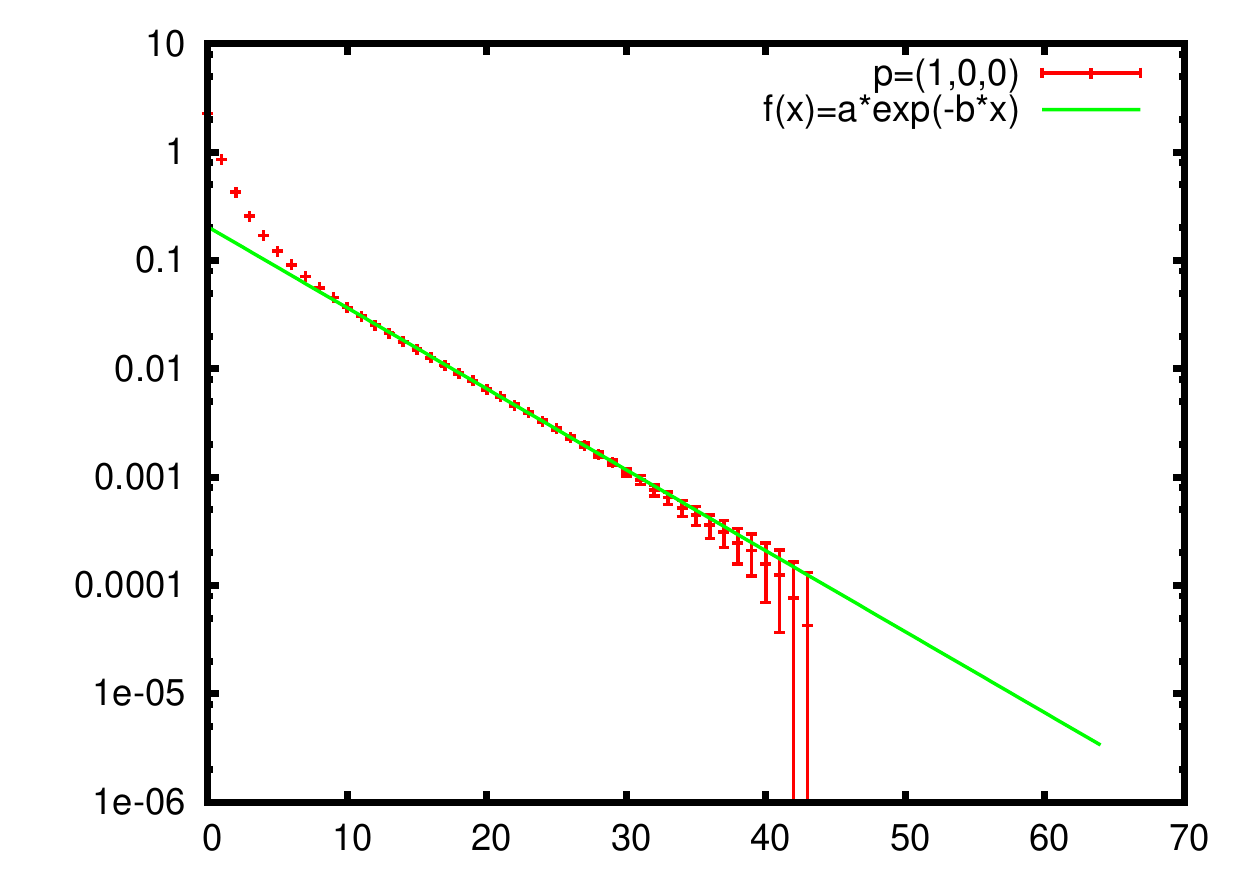}
  \includegraphics[width=.49\textwidth]{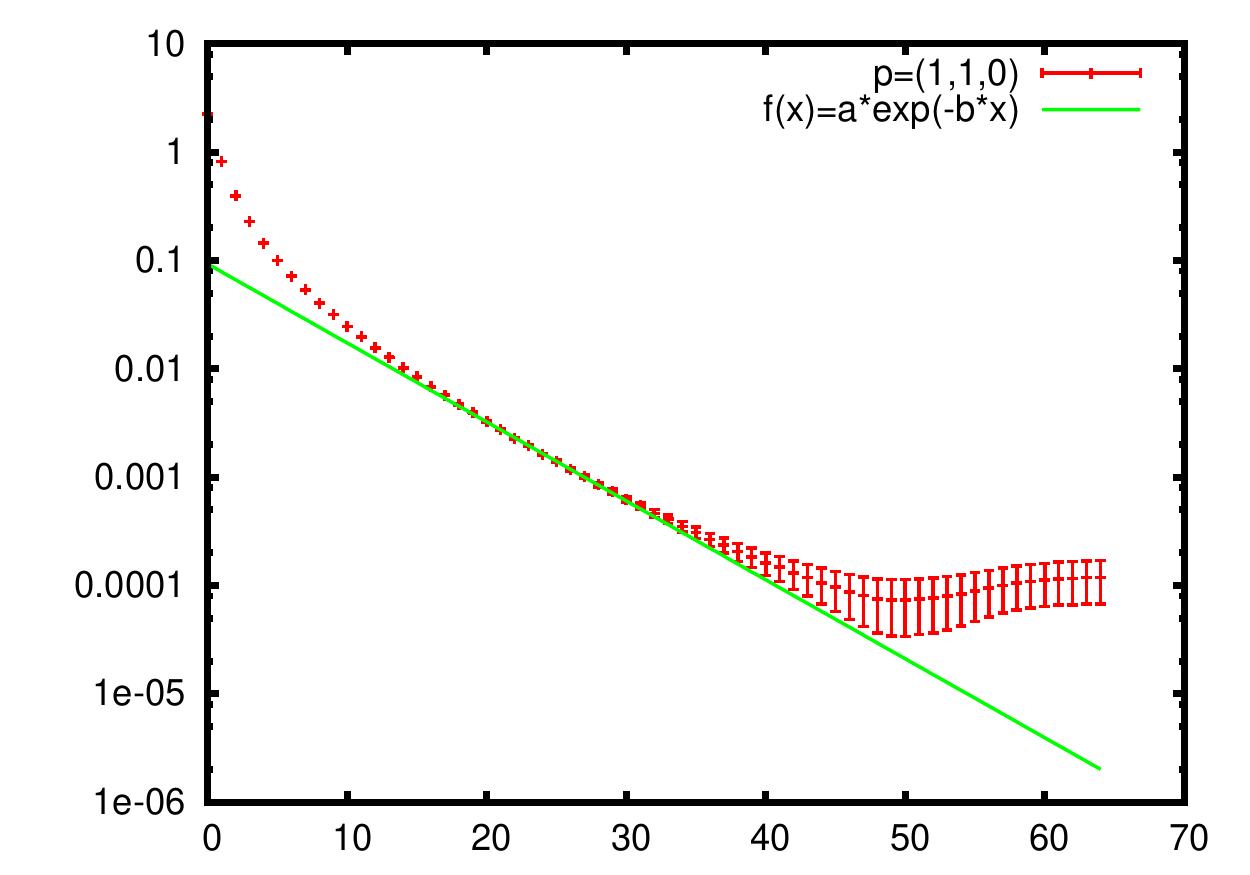}
  \caption{The correlators in Figure 2 for the two lowest
    momentum values with a single exponential fit with uncorrelated
    errors over the range $20:30$.}
  \label{fig3}
\end{figure}

Figure~\ref{fig3} shows single exponential fits to the data with
uncorrelated errors for the $p=1, 0, 0$ component (left panel) and
$p=1,1,0$ component (right panel -- of course, as in Figure 2, all
similar p components are averaged) over the range 20:30.  From the
corresponding coefficients of the exponential fall-offs, $0.1721$ and
$0.1678$ for the lowest and next lowest momenta, one would derive bare
mass values $am=0.1413$ and $am=0.0943$.  The difference between the
two values, the lowest one being in the middle of the error range for
the $\sigma$ mass found in Ref.~\cite{Appelquist:2018yqe}, the other
one a little above the error range, points to the fact that
substantial statistical fluctuations are still present in the data and
validates the caution expressed by the large error quoted in
Ref.~\cite{Appelquist:2018yqe}.  The fits shown in Figure~\ref{fig3}
are rather unsophisticated and should only be given illustrative
value.  The point which we would like to make is that it is rather
easy to calculate the disconnected correlators (as well as the
connected ones, of course) with non-zero momentum and that these can
provide useful information supplementing the information obtained from
zero momentum.  The LSD collaboration is considering including
non-zero momentum correlators in future analyses of the spectrum of
masses.

\section{Conclusions}
\label{conclusions}

Correlators with non-zero momentum in lattice simulations have been
considered many times, so no originality is claimed for this.  What
may be new is the proposal to consider correlators weighted with a
Gaussian distribution that favors the disconnected components at
closer distances\footnote{I will appreciate being informed of any
  previous work that should be cited. If appropriate, I will include
  the citation in the article posted in the archives and, if feasible,
  in the published proceedings.}.  This noise reduction procedure is
of limited or no value for lattices of small to moderate spatial
extension, but may become useful as new, powerful computers permit the
simulation of systems with very large spatial extent.  Because of the
inclusion of non-zero momentum, fits to the data become more
elaborated, but can be done using a fitting function provided by a
subroutine, rather than a combination of exponentials.  Another point
emphasized in this note is that the assumption of momentum
conservation, which becomes better and better satisfied as the
statistics of the simulation increases, allows a simplification in the
calculation of the weighted correlators as well as of the momentum
components of the disconnected correlators.  Correlators with non-zero
momentum avoid the need of a vacuum subtraction and their inclusion in
a global fit may improve the overall accuracy in the extraction of
spectroscopic data from the correlators.

\section*{Acknowledgments}
I am grateful to my colleagues in the LSD collaboration for valuable
discussions and, especially to Enrico Rinaldi and Evan Weinberg, for
providing me with the raw data needed for my analysis of the correlators.

This work was supported by DOE grant DE-SC0015845.

\bibliography{BSM}

%Merlin.mbs v4.21 2009-07-09.
\begin{thebibliography}{1}%
\makeatletter
\providecommand \@ifxundefined [1]{%
 \ifx #1\undefined \expandafter \@firstoftwo
 \else \expandafter \@secondoftwo
\fi
}%
\providecommand \@ifnum [1]{%
 \ifnum #1\expandafter \@firstoftwo
 \else \expandafter \@secondoftwo
\fi
}%
\providecommand \enquote [1]{``#1''}%
\providecommand \bibnamefont  [1]{#1}%
\providecommand \bibfnamefont [1]{#1}%
\providecommand \citenamefont [1]{#1}%
\providecommand\href[0]{\@sanitize\@href}%
\providecommand\@href[1]{\endgroup\@@startlink{#1}\endgroup\@@href}%
\providecommand\@@href[1]{#1\@@endlink}%
\providecommand \@sanitize [0]{\begingroup\catcode`\&12\catcode`\#12\relax}%
\@ifxundefined \pdfoutput {\@firstoftwo}{%
 \@ifnum{\z@=\pdfoutput}{\@firstoftwo}{\@secondoftwo}%
}{%
 \providecommand\@@startlink[1]{\leavevmode\special{html:<a href="#1">}}%
 \providecommand\@@endlink[0]{\special{html:</a>}}%
}{%
 \providecommand\@@startlink[1]{%
  \leavevmode
  \pdfstartlink
   attr{/Border[0 0 1 ]/H/I/C[0 1 1]}%
   user{/Subtype/Link/A<</Type/Action/S/URI/URI(#1)>>}%
  \relax
 }%
 \providecommand\@@endlink[0]{\pdfendlink}%
}%
\providecommand \url  [0]{\begingroup\@sanitize \@url }%
\providecommand \@url [1]{\endgroup\@href {#1}{\urlprefix}}%
\providecommand \urlprefix [0]{URL }%
\providecommand \Eprint[0]{\href }%
\@ifxundefined \urlstyle {%
  \providecommand \doi [1]{doi:\discretionary{}{}{}#1}%
}{%
  \providecommand \doi [0]{doi:\discretionary{}{}{}\begingroup
  \urlstyle{rm}\Url }%
}%
\providecommand \doibase [0]{http://dx.doi.org/}%
\providecommand \Doi[1]{\href{\doibase#1}}%
\providecommand \bibAnnote [3]{%
  \BibitemShut{#1}%
  \begin{quotation}\noindent
    \textsc{Key:}\ #2\\\textsc{Annotation:}\ #3%
  \end{quotation}%
}%
\providecommand \bibAnnoteFile [2]{%
  \IfFileExists{#2}{\bibAnnote {#1} {#2} {\input{#2}}}{}%
}%
\providecommand \typeout [0]{\immediate \write \m@ne }%
\providecommand \selectlanguage [0]{\@gobble}%
\providecommand \bibinfo [0]{\@secondoftwo}%
\providecommand \bibfield [0]{\@secondoftwo}%
\providecommand \translation [1]{[#1]}%
\providecommand \BibitemOpen[0]{}%
\providecommand \bibitemStop [0]{}%
\providecommand \bibitemNoStop [0]{.\EOS\space}%
\providecommand \EOS [0]{\spacefactor3000\relax}%
\providecommand \BibitemShut [1]{\csname bibitem#1\endcsname}%
%</preamble>
\bibitem{Appelquist:2018yqe}%
  \BibitemOpen
  \bibfield{author}{%
  \bibinfo {author} {\bibfnamefont{T.}~\bibnamefont{Appelquist}}, \bibinfo
  {author} {\bibfnamefont{R.}~\bibnamefont{Brower}}, \bibinfo {author}
  {\bibfnamefont{G.}~\bibnamefont{Fleming}}, \bibinfo {author}
  {\bibfnamefont{A.}~\bibnamefont{Gasbarro}}, \bibinfo {author}
  {\bibfnamefont{A.}~\bibnamefont{Hasenfratz}}, \bibinfo {author}
  {\bibfnamefont{X.-Y.}\ \bibnamefont{Jin}}, \bibinfo {author}
  {\bibfnamefont{E.}~\bibnamefont{Neil}}, \bibinfo {author}
  {\bibfnamefont{J.}~\bibnamefont{Osborn}}, \bibinfo {author}
  {\bibfnamefont{C.}~\bibnamefont{Rebbi}}, \bibinfo {author}
  {\bibfnamefont{E.}~\bibnamefont{Rinaldi}}, \bibinfo {author}
  {\bibfnamefont{D.}~\bibnamefont{Schaich}}, \bibinfo {author}
  {\bibfnamefont{P.}~\bibnamefont{Vranas}}, \bibinfo {author}
  {\bibfnamefont{E.}~\bibnamefont{Weinberg}},\ and\ \bibinfo {author}
  {\bibfnamefont{O.}~\bibnamefont{Witzel}} (\bibinfo {collaboration} {Lattice
  Strong Dynamics})}%
   (\bibinfo {year} {2018}),\
  \Eprint{http://arxiv.org/abs/1807.08411}{arXiv:1807.08411 [hep-lat]}%
  \bibAnnoteFile{NoStop}{Appelquist:2018yqe}%
%%CITATION = ARXIV:1807.08411;%%
\end{thebibliography}%
\bibliographystyle{apsrev4-1}

\end{document}